\documentclass[a4paper,11pt]{article}
\usepackage{jcappub}
\usepackage{lineno}

\usepackage{booktabs}
\usepackage[nameinlink,noabbrev]{cleveref}
\usepackage{dcolumn}
\usepackage{physics}
\usepackage{rotating}
\usepackage[separate-uncertainty]{siunitx}
\usepackage{tabularx}

\crefname{equation}{eq.}{eqs.}
\Crefname{equation}{Equation}{Equations}
\DeclareSIUnit{\year}{yr}
\def\TCMB{T_0}
\title{Some Times in Standard Cosmology} 
\author{Lukas Tobias Hergt}
\author{and Douglas Scott}
\affiliation{
    Department of Physics \& Astronomy,
    University of British Columbia,
    Vancouver, BC, V6T\,1Z1, Canada
}
\emailAdd{lthergt@phas.ubc.ca}
\emailAdd{dscott@phas.ubc.ca}
\notoc

\abstract{
    The standard cosmological model is sufficiently well constrained that precise estimates can be provided for the redshift of various physically defined times in the chronology of the Universe.  For example, it is well known that matter-radiation equality, recombination and reionisation happen at redshifts of around 3000, 1000 and 10, respectively, and these can be specified more precisely by fitting to data.  What is less well known are the times in years (and their uncertainties) for these and other epochs in the history of the Universe.  Here we provide precise time determinations for six epochs in cosmological history within the standard model, using data from the \textit{Planck} satellite. Our main results are illustrated in a figure.
}

\begin{document}
\maketitle
\flushbottom

\section{Introduction}

The current standard cosmological model, $\Lambda\mathrm{CDM}$, is defined by seven parameters within a set of simplifying assumption (summarised in \cref{tab:parameters})~\cite{Olive2024,SkepticsGuide}.
Anisotropies in the cosmic microwave background~(CMB) provide the tightest constraints on most of these parameters, through their angular power spectra. Likelihood functions derived from the power spectra, fitted through Markov chain Monte Carlo~(MCMC) techniques, provide sets of samples whose posterior distributions give central values and uncertainties on parameters.
The CMB temperature~$\TCMB$ is distinct from the others, being so well determined that it can be considered to be fixed~\cite{Wen2021}. This leaves us with a 6-dimensional parameter space to determine the best fit to the CMB power spectra.

In addition to the usual six parameters, it is also possible to calculate the constraints on any combination of those parameters, or indeed any function of them
(see e.g.\ Ref.~\cite{Mnemonics} for more examples).
Among the parameters typically discussed are the epochs at which specific physical things happen in the history of the Universe, usually given in terms of the scale factor~$a$ or redshift~$z$. The average non-cosmologist is likely to expect to hear of epochs defined in actual time units, but such values do not occur in most parameter tables. Here we calculate the time in years for several key epochs in cosmic history.

Before starting, it is worth remembering that ``cosmic time'' can be well defined as the time coordinate measured by an observer moving in the Hubble flow (i.e.\ with no peculiar velocity), assuming a homogeneous background.  Hence we can talk meaningfully about time to many digits of precision.

Any epoch can be defined through the use of the scale factor~$a(t)$, or equivalently a redshift~$z(t)$ defined through 
\begin{equation}
    1+z(t)=\frac{1}{a(t)}.
\end{equation}
Within the standard cosmology, the time corresponding to a given scale factor can be found through an integral. If we start with
\begin{equation}
    \label{eq:time}
    t = \int_0^t \dd{t} = \int_0^a \frac{\dd{a}}{\dot{a}} = \int_0^a \frac{\dd{a}}{aH(a)} ,
\end{equation}
then we can use the Friedmann equation in the form
\begin{equation}
    H^2(a) = H_0^2\left\{\Omega_\Lambda + \Omega_\mathrm{m}a^{-3} + \Omega_\mathrm{r}a^{-4}\right\}
\end{equation}
to perform the integral. Here $H\equiv\dot{a}/a$ is the Hubble parameter, quantifying the rate of expansion, and $H_0$ is its value today.
For models with curvature, things are a little more complicated, but we will stick with spatially flat models here (i.e.\ the curvature density parameter is~$\Omega_K=0$).
We are also assuming that the dark energy is a pure cosmological constant~$\Lambda$ (i.e.\ the equation of state parameter is~$w=-1$).

The present-day radiation density parameter~$\Omega_\mathrm{r,0}$ for relativistic species can be computed from the energy densities~$\rho$ for photons~$\gamma$ and neutrinos~$\nu$, together with the monopole temperature $\TCMB$ of the CMB and the effective number of neutrinos~$N_\mathrm{eff}$:
\begin{equation}
\label{eq:Omega_r}
\begin{alignedat}{2}
    \Omega_\mathrm{r,0} &= \frac{\rho_\mathrm{r,0}}{\rho_\mathrm{crit}} = \frac{8\pi G}{3H_0^2} \left(\rho_{\gamma,0}+\rho_{\nu,0}\right) 
    & \quad \text{with} \quad \rho_{\gamma,0} &= \frac{a_\mathrm{B}}{c^2} \TCMB^4 , \\
    && \quad \text{and} \quad \frac{\rho_{\nu,0}}{\rho_{\gamma,0}} &= \frac{7}{8}  N_\mathrm{eff} \left(\frac{4}{11}\right)^{4/3} ,
\end{alignedat}
\end{equation}
and where $G$ is Newton's gravitational constant and $a_\mathrm{B}$ is the radiation constant of the Stefan--Boltzmann law.\footnote{Related to the Stefan--Boltzmann constant $\sigma=c\,a_\mathrm{B}/4$.} Because neutrino decoupling is not instantaneous (as well as some other minor corrections) we set the effective number of neutrinos to $N_\mathrm{eff}=3.0440$ when there are 3 neutrino species~\cite{Bennett2021}.\footnote{Strictly speaking, only relativistic neutrinos should enter into \cref{eq:Omega_r}, which we neglect here.} As such the radiation density today contributes about $\Omega_\mathrm{r,0}h^2\simeq\num{4e-5}$.

In general the time integral from \cref{eq:time} has to be performed numerically, but there are analytic solutions for 2-fluid simplifications, which we will use in \cref{sec:eq,sec:star,sec:q}.
The full numerical integration is already part of so-called Boltzmann solvers that calculate the angular power spectra for perturbations, such as the codes \texttt{CAMB}~\cite{Camb1,Howlett2012} and \texttt{CLASS}~\cite{Class1,Class2,Class3,Class4}.
Hence we can use modified output from such codes to calculate times corresponding to
different redshifts.
The differences between the codes are negligible compared to the statistical uncertainties~\cite{Class3}.

\begin{table}[tbp]
    \centering
    \renewcommand{\arraystretch}{1.25}
    \begin{tabularx}{\textwidth}{ l c r@{${}\pm{}$}l }
        \toprule
\multicolumn{1}{X}{Description}                           & Parameter                  & \multicolumn{2}{c}{Value} \\
        \midrule
Baryon density today                                      & $\Omega_\mathrm{b,0}h^2$   &   $0.02225$ & $0.00013$   \\
Cold dark matter density today                            & $\Omega_\mathrm{c,0}h^2$   &    $0.1190$ & $0.0011$    \\
Acoustic angular scale at last scattering (approximation) & $100\,\theta_\mathrm{MC}$  &   $1.04085$ & $0.00026$   \\
Thomson scattering optical depth due to reionisation      & $\tau_\mathrm{reio}$       &     $0.059$ & $0.006$     \\
Log-amplitude of primordial density perturbations         & $\ln(10^{10}A_\mathrm{s})$ &     $3.045$ & $0.012$     \\
Scalar spectral index of primordial density perturbations & $n_\mathrm{s}$             &     $0.968$ & $0.004$     \\
        \midrule
CMB monopole temperature today                            & $\TCMB$                    & \multicolumn{2}{c}{\SI{2.7255}{\K}} \\
Curvature density parameter today (assuming flatness)     & $\Omega_K$                 & \multicolumn{2}{c}{0} \\
Sum of neutrino masses assuming 1 massive neutrino        & $\sum m_\nu$               & \multicolumn{2}{c}{\SI{0.06}{\eV}} \\
Effective number of massive and massless neutrinos        & $N_\mathrm{eff}$           & \multicolumn{2}{c}{3.0440} \\
Equation of state parameter for dark energy               & $w$                        & \multicolumn{2}{c}{$-1$} \\
        \bottomrule
    \end{tabularx}
    \caption{
        The six cosmological sampling parameters (upper block; showing sample mean and standard deviation), and some fixed parameters (lower block) from the MCMC run with the \textit{Planck} likelihoods \texttt{lowlTT+lowlEE+highlTTTEEE+lensing} (see \cref{sec:methods} for technical details). 
    }
    \label{tab:parameters}
\end{table}

In the rest of this paper we will consider several specific epochs in the history of the Universe within the context of the standard $\Lambda$CDM cosmology.  We will specifically focus on the time of: matter-radiation equality; recombination; baryon-temperature decoupling; reionisation; the transition from deceleration to acceleration; and the present day.

\section{Methods and data}
\label{sec:methods}

For the computation of the angular power spectra of temperature and polarisation anisotropies, and for various derived parameters we use the cosmological Boltzmann code \texttt{CAMB}~\cite{Camb1,Howlett2012}. The posterior distribution has been explored using \texttt{Cobaya}~\cite{Cobaya1}, linking the Markov chain Monte Carlo~(MCMC) sampler~\cite{CosmoMC1,CosmoMC2,CosmoMC3} with the CMB likelihoods from \textit{Planck}. 

For the low-$\ell$ temperature likelihood we used the \texttt{Python} translation by Eirik Gjerløw (Feb.~2023) of the Planck~2018 (also referred to as Public Release 3 or PR3) Gibbs $TT$ likelihood~\cite{Planck2018Likelihoods}.
We also use the low-$\ell$ $E$-mode polarisation likelihood called \texttt{LoLLiPoP}~\cite{Tristram2021,Mangilli2015,Hamimeche2008} and the high-$\ell$ $TT,TE,EE$ cross-spectra likelihood called \texttt{HiLLiPoP}~\cite{Tristram2024}, which both make use of the re-analysed \textit{Planck} maps from the Public Release~4~(PR4), also called \texttt{NPIPE}~\cite{NPIPE}.
Finally, we use the update to the CMB lensing likelihood, which also used PR4 data~\cite{Carron2022,Carron2022a}. Together, this set of likelihoods currently provides the most stringent constraints on cosmological parameters from CMB data alone, and we refer to this combination of likelihoods in this paper as \texttt{lowlTT+lowlEE+highlTTTEEE+lensing}.

Note that in all the results that follow, we compute redshifts and times as derived parameters within the MCMC chains, rather than just converting the mean and the error bar of the redshift to a corresponding time value. As such, we obtain their full, joint posterior distribution.

\section{Results}
\label{sec:results}

\begin{figure}[tp]
    \centering
    \includegraphics[width=\linewidth]{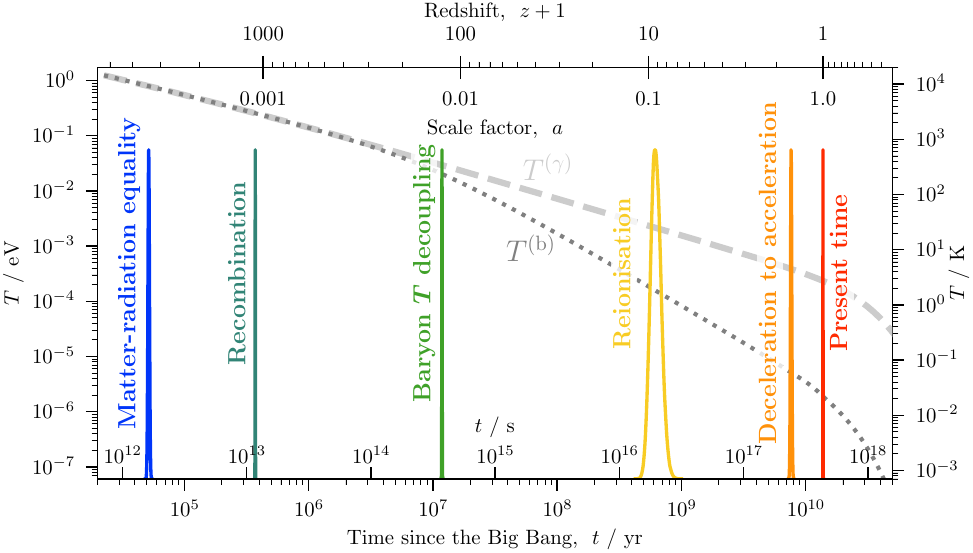}
    \caption{
        Representation of cosmic chronology.  We plot the full one-dimensional marginalised posterior distributions from the MCMC run for the epochs of matter-radiation equality~$t_\mathrm{eq}$, recombination~$t_\ast$, baryon-temperature decoupling~$t_T$, reionisation~$t_\mathrm{reio}$, deceleration to acceleration~$t_q$, and the age of the Universe~$t_0$. 
        The fact that these are posterior distributions (normalised to share the same height) is best seen for the epoch of reionisation; the other cases appear almost as vertical lines due to their small uncertainties when plotted on the large range of the time axis, a testament to the incredible precision that CMB data provide us with. 
        The posteriors are plotted with respect to the time axis at the bottom and the secondary (top) axes for redshift and scale factor are scaled relative to the time axis by computing the cosmology from the average values of the cosmological sampling parameters.
        Since the Universe is dominated by matter for almost all of the time window shown here, the log-redshift and log-scale-factor scale linearly with the log-time. Only towards the end of the time window, past the present time, does the effect of $\Lambda$ domination taking over become visible in the form of the shrinking minor ticks of the redshift axis. 
        The time dependence of the photon temperature~$T^{(\gamma)}$ and baryon temperature~$T^\mathrm{(b)}$ (plotted as grey dashed and grey dotted lines, respectively) make this even clearer. With respect to the scale factor~$a$, the photon temperature follows a perfect power law, and the baryons depart from this at baryon decoupling.  With respect to physical time~$t$, both $T^{(\gamma)}$ and $T^\mathrm{(b)}$ start cooling much faster once dark energy takes over and the expansion of the Universe starts to accelerate. Note that the baryon temperature increases around the epoch of reionisation, but we have only plotted its evolution in a completely homogeneous model.
    }
    \label{fig:times}
\end{figure}

The various times, redshifts, scale factors, and temperatures associated with the already mentioned epochs are listed in \cref{tab:times}, and \cref{fig:times} connects all of them in a single plot, highlighting how each of these quantities allow us to track the passing of time (in the general sense, here) in our expanding Universe.  We focus on six specific epochs, which we describe in turn below.

\begin{table}[tbp]
    \centering
    \begin{tabularx}{\textwidth}{ l X r@{${}\pm{}$}l r@{${}\pm{}$}l }
\toprule
Cosmic (physical) time & $t_\mathrm{eq}$              & $(51.7$    & $0.8)~\mathrm{kyr}$   & $(1.633$   & $0.024)\times10^{12}~\mathrm{s}$ \\
                       & $t_\ast$                     & $(372.6$   & $1.0)~\mathrm{kyr}$   & $(1.1759$  & $0.0030)\times10^{13}~\mathrm{s}$ \\
                       & $t_T$                        & $(11.86$   & $0.04)~\mathrm{Myr}$  & $(3.743$   & $0.013)\times10^{14}~\mathrm{s}$ \\
                       & $t_\mathrm{reio}$            & $(630$     & $60)~\mathrm{Myr}$    & $(1.98$    & $0.20)\times10^{16}~\mathrm{s}$ \\
                       & $t_q$                        & $(7.66$    & $0.09)~\mathrm{Gyr}$  & $(2.416$   & $0.029)\times10^{17}~\mathrm{s}$ \\
                       & $t_0$                        & $(13.808$  & $0.020)~\mathrm{Gyr}$ & $(4.358$   & $0.006)\times10^{17}~\mathrm{s}$\\ \addlinespace
Redshift               & $z_\mathrm{eq}$              & $3377$     & $25$                  \\
                       & $z_\ast$                     & $1090.00$  & $0.23$                \\
                       & $z_T$                        & $124.07$   & $0.26$                \\
                       & $z_\mathrm{reio}$            & $8.2$      & $0.6$                 \\
                       & $z_q$                        & $0.642$    & $0.017$               \\
                       & $z_0$                        & \multicolumn{2}{c@{\hphantom{$\pm$}}}{$0$\hphantom{$\times10^{-4}$}} \\ \addlinespace
Scale factor           & $a_\mathrm{eq}$              & $(2.960$   & $0.021)\times10^{-4}$ \\
                       & $a_\ast$                     & $(9.1659$  & $0.0019)\times10^{-4}$ \\
                       & $a_T$                        & $(7.995$   & $0.017)\times10^{-3}$ \\
                       & $a_\mathrm{reio}$            & $0.110$    & $0.007$               \\
                       & $a_q$                        & $0.609$    & $0.006$               \\
                       & $a_0$                        & \multicolumn{2}{c@{\hphantom{$\pm$}}}{$1$\hphantom{$\times10^{-4}$}} \\ \addlinespace
Photon temperature     & $T^{(\gamma)}_\mathrm{eq}$   & $(9210$    & $70)~\mathrm{K}$      & $(0.793$   & $0.006)~\mathrm{eV}$  \\
                       & $T^{(\gamma)}_\ast$          & $(2973.5$  & $0.6)~\mathrm{K}$     & $(0.25624$ & $0.00005)~\mathrm{eV}$ \\
                       & $T^{(\gamma)}_T$             & $(340.9$   & $0.7)~\mathrm{K}$     & $(29.38$   & $0.06)~\mathrm{meV}$  \\
                       & $T^{(\gamma)}_\mathrm{reio}$ & $(24.9$    & $1.6)~\mathrm{K}$     & $(2.15$    & $0.14)~\mathrm{meV}$  \\
                       & $T^{(\gamma)}_q$             & $(4.48$    & $0.05)~\mathrm{K}$    & $(0.386$   & $0.004)~\mathrm{meV}$ \\
                       & $T^{(\gamma)}_0$             & \multicolumn{2}{c@{\hphantom{$\pm$}}}{$2.7255~\mathrm{K}$\hphantom{$\times10^{-4}$}}
                                                      & \multicolumn{2}{c@{\hphantom{$\pm$}}}{$0.23487~\mathrm{meV}$\hphantom{$\times10^{-4}$}} \\ \addlinespace
\bottomrule
\end{tabularx}
    \caption{
        Posterior average and standard deviation of the derived time~$t$, redshift~$z$, scale factor~$a$, and photon temperature~$\smash{T^{(\gamma)}}$ for the epochs of matter-radiation equality\,($_\mathrm{eq}$), recombination\,($_\ast$), baryon-temperature decoupling\,($_T$), reionisation\,($_\mathrm{reio}$), deceleration to acceleration\,($_q$), and for today\,($_0$). Note that the present-day redshift~$z_0$ and scale factor~$a_0$ are fixed by definition, and the CMB monopole temperature~$\smash{T^{(\gamma)}_0=\TCMB}$ is set to match the high-precision measurement by the FIRAS instrument on the \textit{COBE} satellite~\cite{FIRAS,Fixsen2009}.
    }
    \label{tab:times}
\end{table}

\subsection{Matter-radiation equality}
\label{sec:eq}

We can calculate the time at which matter and radiation (including the neutrino contribution) densities are equal, which is commonly labelled with a subscript ``eq'', such as for the redshift~$z_\mathrm{eq}$, which is one of the derived parameters reported in the \textit{Planck} parameter tables~\cite{Planck2013Parameters,Planck2015Parameters,Planck2018Parameters}.
Because the relevant 2-fluid solution is analytic, there is an explicit expression for the corresponding time:
\begin{equation}
    \label{eq:t_eq}
    t_\mathrm{eq} = \frac{4\,a_\mathrm{eq}^2}{3H_0\sqrt{\Omega_\mathrm{r,0}}} \left(1-\frac{1}{\sqrt{2}}\right).
\end{equation}
We find $z_\mathrm{eq}=\num{3377(25)}$ and $t_\mathrm{eq}=\SI{51.7(8)}{\kilo\year}$ using the full numerical solutions from \texttt{CAMB}. The maximum deviation of the approximation in \cref{eq:t_eq} is $\lesssim\SI{0.002}{\kilo\year}$ (corresponding to relative deviation of about \num{3e-5}), and thus far smaller than the statistical uncertainty.

\subsection{Recombination}
\label{sec:star}

The recombination epoch provides us with a whole set of parameters, such as $100\,\theta_\ast=\SI{1.04106(25)}{}$, the angular size of the sound horizon at last scattering, the best constrained of the usual six sampling parameters.\footnote{Note that for MCMC runs with \texttt{CAMB}, rather than sampling $\theta_\ast$ directly, it is approximated with the sampling parameter $\theta_\mathrm{MC}$. The actual $\theta_\ast$ is turned into a derived parameter that almost matches $\theta_\mathrm{MC}$, but not quite. Also, there are different points of reference that can be chosen for the epoch of recombination. Two common ones are the maximum of the visibility function, in which case quantities are typically labelled with a subscript~``rec'', or the point where the Thomson optical depth crosses one, typically labelled with a subscript asterisk~($_\ast$). We focus on the latter here.} 
The redshift of the recombination epoch $z_\ast\simeq1100$ is an often quoted number, physically defined as the redshift, integrated back from today, where the Thomson-scattering optical depth reaches unity (assuming for this calculation that there is no reionisation). 
We confirm the value to be $z_\ast=\num{1090.00(23)}$.
The corresponding time can still be approximated by the analytic expression of the 2-fluid matter-radiation solution (with a maximum deviation of about \SI{0.03}{\kilo\year}, corresponding to a relative deviation of about \num{9e-5}):
\begin{equation}
    \label{eq:t_star}
    t_\ast = \frac{4 a_\mathrm{eq}^2}{3H_0\sqrt{\Omega_\mathrm{r,0}}} \left[1- \left(1-\frac{a_\ast}{2a_\mathrm{eq}}\right) \left(1+\frac{a_\ast}{a_\mathrm{eq}}\right)^{1/2} \right].
\end{equation}
We find $t_\ast=\SI{372.6(10)}{\kilo\year}$.

\subsection{Baryon-temperature decoupling}
\label{sec:T}

One further time we can calculate is the decoupling time for baryons,\footnote{Many authors use the ambiguous term ``decoupling'' to refer to recombination. However, the huge photon-to-baryon ratio means that baryons decouple from photons much later than photons decouple from baryons.} i.e.\ the time when the atoms stopped being held at the CMB temperature by Compton scattering and were able to cool more rapidly~\cite{Scott2009}. 

We define the time of baryon-temperature\footnote{This is sometimes more generally referred to as the ``matter temperature'', as it is technically the electrons that the photons interact with, and then protons and atoms are coupled with them through other scattering processes.} decoupling~$t_T$ as the intersection point between the limiting behaviour of the photon and baryon temperature scaling after decoupling. The photons continue to scale as $T^{(\gamma)}\propto a^{-1}$, but the baryons starts scaling as $T^\mathrm{(b)}\propto a^{-2}$. Using today's photon temperature~$T^{(\gamma)}_0$ and baryon temperature~$T^\mathrm{(b)}_0$ (ignoring heating effects and reionisation for this specific calculation), we can infer the intersection point:
\begin{align}
    1+z_T = \frac{1}{a_T} = \frac{T^{(\gamma)}_0}{T^\mathrm{(b)}_0}.
\end{align}
We can still use the approximation from \cref{eq:t_star} for the conversion from redshift to cosmic time. While the maximum relative deviation of this approximation increases to about \num{7e-4}, since this is roughly half way to the point where $\Lambda$ domination will take over; with a maximum shift of about \SI{0.008}{\mega\year} this is still well below the statistical uncertainty.

We find $z_T=\num{124.07(26)}$ and $t_T=\SI{11.86(04)}{\mega\year}$. After this epoch, the baryonic matter is able to depart from the photon bath, cooling adiabatically as $T^\mathrm{(b)}\propto(1+z)^2$.  Hence today, if there had been no heating, then we would have $T^\mathrm{(b)}_0\simeq\SI{22}{\milli\kelvin}$.
In reality, it is expected that at some time after $t_T$, shock-heating from structure formation raises the gas temperature well above that of the CMB \cite{SunZel1972,MossScott2009}, and then energy injection from photons adds further heating sources later, the details of which are not yet known \cite{BarkanaLoeb01,Furlanetto2006}.

\subsection{Reionisation}
\label{sec:reio}

For more recent epochs, we can use the solution for a flat model consisting of non-relativistic matter and a cosmological constant. With a small present-day radiation density parameter of $\Omega_\mathrm{r,0}\lesssim\num{e-4}$ (see also \cref{eq:Omega_r}), such a 2-fluid assumption is accurate to four digits at low redshift, when radiation can be neglected. The expression is
\begin{equation}
    \label{eq:t_q}
    t = \frac{2}{3H_0\sqrt{\Omega_{\Lambda,0}}} \sinh^{-1}\left(\sqrt{\frac{\Omega_{\Lambda,0}}{(1-\Omega_{\Lambda,0})(1+z)^3}}\right).
\end{equation}

The formation of structure at relatively low redshifts leads to reionisation of the largely neutral matter in the Universe.  This transition could have many functional forms, but for existing data the effects on the CMB power spectra come mainly from the overall
optical depth~$\tau_\mathrm{reio}$, with very little sensitivity to the shape of the transition. Hence we can focus on the parameter $\tau_\mathrm{reio}$ and assume for simplicity that the reionisation happens rapidly at redshift $z_\mathrm{reio}$, or equivalently we can define $z_\mathrm{reio}$ as the redshift  when the Universe is half ionised.\footnote{In more detail, the codes require some non-zero width for the reionisation process, using $\Delta z=0.5$ for example. However, these details make little quantitative difference.}

We find $z_\mathrm{reio}=\num{8.2(6)}$, corresponding to $t_\mathrm{reio}=\SI{630(30)}{\mega\year}$.
With a maximum deviation of $\lesssim\SI{2.4}{\mega\year}$ (corresponding to a relative deviation of about \num{4e-3}), \cref{eq:t_q} is a less accurate approximation than \cref{eq:t_star} for the previous epochs, but it is still well below the statistical uncertainty.

\subsection{Deceleration to acceleration}
\label{sec:q}

The epoch at which the Universe transitioned from deceleration to
acceleration is given by
\begin{equation}
    1+z_q = \left( \frac{2\,\Omega_{\Lambda,0}}{\Omega_\mathrm{m,0}} \right)^{1/3}
\end{equation}
and the CMB data yield $z_q=\num{0.642(17)}$.
For the corresponding time, we find $t_q=\SI{7.66(9)}{\giga\year}$.
Approximating with \cref{eq:t_q} leads to a maximal deviation of $\lesssim\SI{0.005}{\giga\year}$, corresponding to a relative deviation of about \num{7e-4}.

\subsection{Today}
\label{sec:0}

We end with the estimate of the age of the Universe today, which is of course well known.  With the current data set we obtain the value $t_0=\SI{13.808(20)}{\giga\year}$, an uncertainty of only about \SI{0.14}{\percent}.

\section{Conclusions}

\begin{figure}[tbp]
    \centering
    \includegraphics[width=\linewidth]{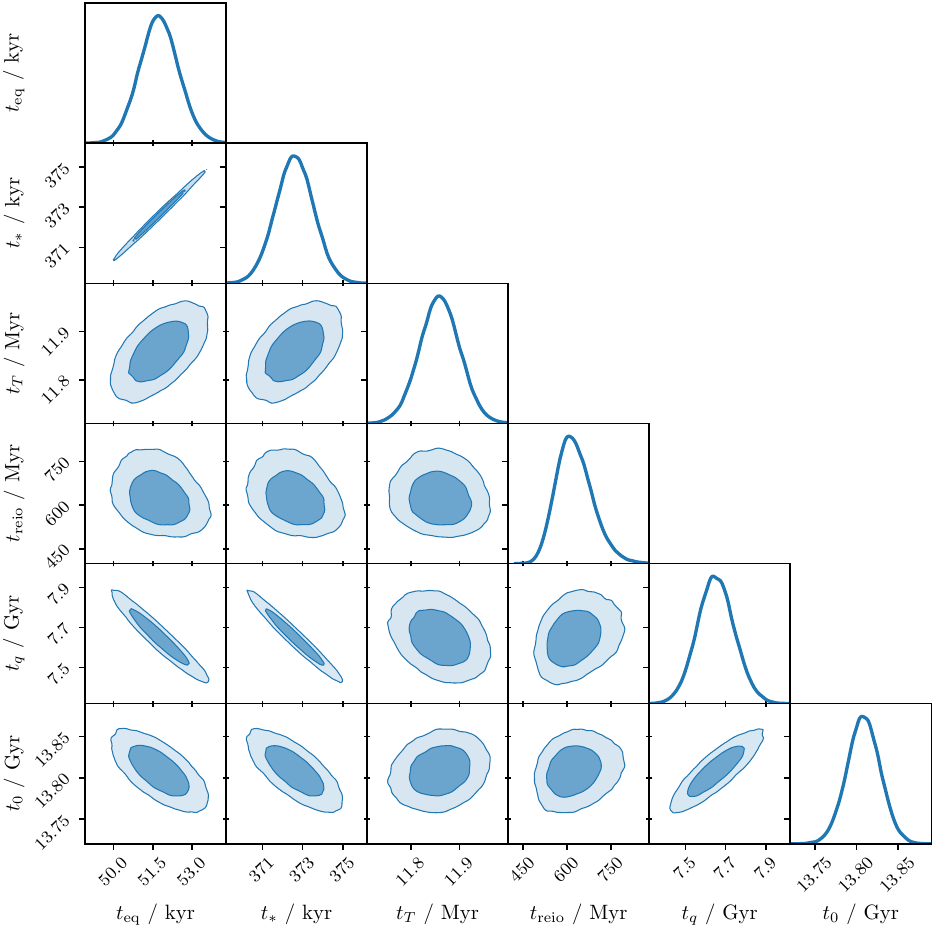}
    \caption{
        Correlations between the time parameters for matter-radiation equality~$t_\mathrm{eq}$, recombination~$t_\ast$, baryon-temperature decoupling~$t_T$, reionisation~$t_\mathrm{reio}$, deceleration to acceleration~$t_q$, and the age of the universe~$t_0$.
    }
    \label{fig:times_2d}
\end{figure}

\Cref{fig:times_2d} shows the joint posteriors of the epochs, where we can clearly see that some of them are correlated with each other.  For example the recombination and equality times have a very strong positive correlation, while the deceleration time is anti-correlated with both
recombination time and equality time.
Baryon-temperature decoupling and reionisation, on the other hand, are comparatively uncorrelated with all other times.

\Cref{fig:times} is the main summary of this paper. There have been many earlier figures describing the chronology of our Universe, notable examples include: the ``Possible Thermal History of the Universe'' by Bob Dicke and collaborators~\cite{DPRW1965}; diagrams for different cosmologies made by Martin Rees~\cite{Rees1981}; the ``Complete history of the Universe'' poster made by Michael Turner and Angela Gonzalez~\cite{Turner2022}; and ``Important Events in the History and Future of the Universe'' by Fred Adams and Greg Laughlin~\cite{AdamsLaughlin1997}.  A major difference with our figure is that now the model is known sufficiently well that precise uncertainties can be attached to the epochs.

Other special periods in cosmic history could also have been considered, for example baryon drag, helium reionisation, matter--dark-energy equality ($\Omega_\textrm{m}(z)=\Omega_\Lambda(z)$), specific times related to Big Bang nucleosynthesis, high-energy particle physics, etc. However, mostly these are either closely related to the six epochs that we focused on, or they are not simple to define.

Data from the current generation of CMB experiments has enabled us to pin down the standard cosmological model to a sufficient extent that key epochs in the history of the Universe can now be determined with precise values and uncertainties. As well as giving redshifts for these epochs, it can also be done in normal every-day time units (e.g.\ years), as we have presented here.

\section*{Data and code availability}
The MCMC chains and plotting scripts are available on Zenodo~\cite{zenodo}:
\newline\url{https://doi.org/10.5281/zenodo.14054334}

\acknowledgments
This research was enabled in part by support provided by the Digital Research Alliance of Canada~(\href{https://alliancecan.ca}{alliancecan.ca}).
LTH was supported by a Killam Postdoctoral Fellowship and a CITA National Fellowship.

\newpage
\bibliographystyle{JHEP}
\bibliography{references}

\end{document}